\documentclass[aip,reprint,amsmath,amssymb]{revtex4-2}

\usepackage{amsmath}    
\usepackage{dcolumn}   
\usepackage{graphicx}  
\usepackage{bm}        
\usepackage{amssymb}   

\usepackage{color}

\newcommand{\tens}[1]{\underline{\underline{#1}}}
\newcommand{\xc}{\mathrm{xc}}

\begin{document}

\title{Self-interaction correction schemes for non-collinear spin-density-functional theory}

\author{Nicolas Tancogne-Dejean}
  \email{nicolas.tancogne-dejean@mpsd.mpg.de}
 \affiliation{Max Planck Institute for the Structure and Dynamics of Matter and Center for Free-Electron Laser Science, Luruper Chaussee 149, 22761 Hamburg, Germany}
 \affiliation{European Theoretical Spectroscopy Facility (ETSF)}

\author{Martin L\"uders}
 \affiliation{Max Planck Institute for the Structure and Dynamics of Matter and Center for Free-Electron Laser Science, Luruper Chaussee 149, 22761 Hamburg, Germany}

 \author{Carsten A. Ullrich}
 \affiliation{Department of Physics and Astronomy, University of Missouri, Columbia, Missouri 65211, USA}

\begin{abstract}
We extend some of the well established self-interaction correction (SIC) schemes of density-functional theory to the case of systems with noncollinear magnetism. Our proposed SIC schemes are tested on a set of molecules and metallic clusters in combination with the widely used local spin-density approximation. As expected from the collinear SIC, we show that the averaged-density SIC works well for improving ionization energies but fails to improve more subtle quantities like the dipole moments of polar molecules. We investigate the exchange-correlation magnetic field produced by our extension of the Perdew-Zunger SIC, showing that it is not aligned with the local total magnetization, thus producing an exchange-correlation torque.
\end{abstract}

\maketitle

\section{Introduction}

In practical (spin) density-functional theory (DFT) calculations, one needs to select an approximate functional of the density to compute the exchange-correlation energy and the corresponding potential.\cite{Perdew2012} Most of the commonly employed approximations are known to suffer from the so-called self-interaction error,\cite{PhysRevB.23.5048} an error that implies that the electron can interact with itself via the total electronic density. The self-interaction error can lead to problems in the prediction of the electronic properties of molecules and materials. For example, it can cause an underestimation of the bandgap of insulators and semiconductors, and an underestimation of the ionization potential and electron affinity of molecules. Thus, correcting for the self-interaction error is important for obtaining reliable DFT predictions of the electronic properties of molecules and materials.\cite{Cohen2012}

The search for schemes correcting the self-interaction error, known as self-interaction correction (SIC), has been pioneered by Perdew and Zunger.\cite{PhysRevB.23.5048} Their proposed method, now referred to as the Perdew-Zunger self-interaction correction (PZ-SIC), leads to an exchange-correlation energy functional that is an explicit functional
of the orbitals and, hence, an implicit density functional. Implementations of the PZ-SIC approach are often done in a generalized Kohn-Sham sense,\cite{Seidl1996}
where the exchange-correlation potential depends on the orbital on which it acts. Alternatively, and in the spirit of the original Kohn-Sham DFT, a
local multiplicative exchange-correlation potential can be constructed from PZ-SIC using the optimized effective potential (OEP) technique.\cite{Talman1976}
The so constructed exchange-correlation potentials have the correct asymptotic behavior and exhibit discontinuities as a function of particle number.\cite{Cohen2012,Vieira2009}

It is possible to solve the OEP equations exactly,\cite{RevModPhys.80.3,Yang2016} but this is known to be numerically challenging, and one often resorts to the scheme introduced by Krieger, Li, and Iafrate (KLI) to approximate the full solution of the OEP equations.\cite{Krieger1992}
A further simplification of the KLI approach is the Slater approximation, which neglects the orbital-dependent part in the OEP equations and replaces it by an orbital-averaged term.\cite{Krieger1992} The so-called globally averaged method (GAM) is defined in a similar spirit.\cite{PhysRevA.58.367,PhysRevA.62.053202}
An even more drastic approximation for the SIC consists in replacing in the PZ-SIC the orbital-dependent part directly by an averaged value for all orbitals, leading to the average-density SIC (AD-SIC).\cite{legrand2002comparison}  More recently, Perdew and coworkers proposed new schemes like the local-scaling SIC~\cite{zope2019step} which are intended to fix some of the known deficiencies of the original PZ-SIC.

To our knowledge, all of these methods have so far only be proposed and employed in the context of collinear spin DFT (SDFT).
However, there exist many electronic systems in which noncollinear magnetism, spin-orbit coupling (SOC) and other relativistic effects are relevant, and often the DFT practitioners are left with no other option than to use the local spin density approximation (LSDA), which suffers from self-interaction error. It is the goal of this paper to explore how to extend the applicability of the above mentioned SIC schemes to the realm of noncollinear magnetism.\cite{PhysRevB.13.4274,PhysRevB.81.125114,PhysRevB.98.035140}
This allows one to include effects stemming from the noncollinear magnetism and at the same time improve upon the LSDA.

Extending the existing SIC schemes to treat noncollinear magnetism requires care: important fundamental conditions are the
local SU(2) gauge invariance of the exchange-correlation energy, and the requirement that the method properly reduces to the collinear limit.
Moreover, an important question is whether the exchange-correlation magnetic field produced by a noncollinear SIC can exert a local torque on the magnetization.\cite{Pluhar2019,Hill2023}
If such a torque exists, it must satisfy the condition that the system cannot exert a global torque on itself
(this is known as the zero-torque theorem of SDFT).\cite{PhysRevLett.87.206403}
It is the goal of this work to discuss these points.

The paper is organized as follows. In Sec.~\ref{sec:theory}, we present the motivation underlying our proposed SIC and extend the collinear formulation of PZ-SIC and AD-SIC to the noncollinear case. Then, in Sec.~\ref{sec:results} we report numerical results obtained for several isolated systems, for which we analyze the effect of the SIC on the electronic and magnetic properties of atoms, small molecules, and clusters. We also discuss its effect on the local texture of the exchange-correction torque. Finally we draw our conclusions in Sec.~\ref{sec:discussion}.

\section{Theory}
\label{sec:theory}

We begin by defining the concept of self-interaction for the general case of noncollinear spin systems. Self-interaction is usually introduced
separately for exchange and correlation. Thus, let us first consider
the exact exchange energy of a system of $N$ electrons,~\cite{PhysRevB.107.165111}
\begin{equation}
 E_{\mathrm{x}}[n,\mathbf{m}] = - \frac{1}{2} \int \int \frac{d\mathbf{r}d\mathbf{r'}}{|\mathbf{r}-\mathbf{r'}|} \mathrm{Tr}\Big[\tens{\gamma}(\mathbf{r},\mathbf{r'}) \tens{\gamma}(\mathbf{r'},\mathbf{r}) \Big] \,,
 \label{eq:energy_exchange_hole}
\end{equation}
where $\mathrm{Tr}$ is the trace over spin indices of the one-particle spin density matrix
$\gamma_{\sigma\tau}(\mathbf{r},\mathbf{r'}) = \sum_j^N \psi_{j\sigma}(\mathbf{r})\psi^*_{j\tau}(\mathbf{r'})$,
constructed from two-component
spinor Kohn-Sham orbitals, where $\sigma = \uparrow,\downarrow$ and likewise for $\tau$.
Here, the double underline in $\tens{\gamma}$ represents a  $2\times 2$ matrix in spin space.\cite{PhysRevB.98.035140}
The Hartree energy is given by
\begin{equation}
 E_{\mathrm{H}}[n] = \frac{1}{2} \int \int d\mathbf{r}d\mathbf{r'}\frac{n(\mathbf{r})n(\mathbf{r'})}{|\mathbf{r}-\mathbf{r'}|}  \,,
 \label{eq:energy_hartree}
\end{equation}
where $n(\mathbf{r}) = \mathrm{Tr}[\tens{\gamma}(\mathbf{r},\mathbf{r})]$ is the total charge density of the system.

From the above definitions of $E_{\mathrm{x}}$ and $E_{\mathrm{H}}$, it is straightforward to show that in the one-electron case we have
\begin{equation}
 E_{\mathrm{x}}[n_i,\mathbf{m}_i] + E_{\mathrm{H}}[n_i] = 0\,,
 \label{eq:self_interaction}
\end{equation}
where $n_i$ and $\mathbf{m}_i$ are the single orbital charge and magnetization densities. This is the generalization of the result shown in Ref.~\onlinecite{PhysRevB.23.5048} for the collinear case, and forms the basis of the self-interaction corrections that we are proposing below.

More generally, for a single orbital there is no correlation energy, so we can write that the exchange-correlation energy should fulfill the constraint
\begin{equation}
 E_{\mathrm{xc}}[n_i,\mathbf{m}_i] + E_{\mathrm{H}}[n_i] = 0\,.
 \label{eq:self_interaction_xc}
\end{equation}
Importantly, we remark here that both the exchange energy, Eq. (\ref{eq:energy_exchange_hole}), and the Hartree energy, Eq. (\ref{eq:energy_hartree}), are invariant under local SU(2) rotations of the spin. We thus obtain from Eq. (\ref{eq:self_interaction_xc}) that the property remains true 
if we rotate the orbitals such that their magnetization aligns with the $z$ direction:
\begin{equation}
 E_{\mathrm{xc}}[n_i, \hat R_z{\bf m}_i] + E_{\mathrm{H}}[n_i] = 0\,,
\end{equation}
where $\hat R_z{\bf m}_i$ is a symbolic operator notation for performing a rotation on the spin parts of all orbitals such that they are reckoned with respect to a given global $z$-axis, and then constructing the resulting orbital magnetizations.

This allows us to make the link with the collinear result, see Eq. (30) of Ref.~\onlinecite{PhysRevB.23.5048}. Of course, when starting from the noncollinear formulation of SDFT, one needs to break some symmetries to reduce the four-component noncollinear theory based on the variables ($n,\mathbf{m}$) into a two-component collinear theory based on the variables ($n, m_z$). This can be achieved for instance using a uniform magnetic field of small magnitude along the $z$-axis, which causes the orbitals to align their magnetization along this direction. In other words, the system needs to be told to choose the $z$-axis as its spin quantization axis.

From this, we obtain a set of necessary conditions to be able to employ Eq. (\ref{eq:self_interaction}) to build a SIC.
The first condition is that the approximate exchange-correlation functional must be locally SU(2) gauge invariant, i.e.,
it produces the same exchange-correlation energy independently of the orientation of the orbital magnetization.

The second condition is that the noncollinear and collinear functionals should produce the same energy for the same density, for a magnetization along the $z$ direction.
In other words, 
$E^{\rm noncoll.}_{\mathrm{xc}}[n_i, m_{i,z}\mathbf{\hat{e}_z}] = E^{\rm coll.}_{\mathrm{xc}}[n_{i\sigma}, 0]$, 
where it is stipulated that $m_{i,z}=n_{i,\uparrow\uparrow} - n_{i,\downarrow\downarrow}$
and $n_{i,\downarrow\uparrow} = n_{i,\uparrow\downarrow} =  0$ (and hence $m_{i,x} = m_{i,y} = 0$).
The collinear functional $E^{\rm coll.}_{\mathrm{xc}}[n_{i\sigma}, 0]$ appears in the definition of PZ-SIC, see below.

These conditions are naturally fulfilled by the LSDA when using the method proposed originally by K\"ubler \textit{et al.}\cite{kubler1988density} The first condition is also
fulfilled by the more recently proposed noncollinear exchange meta-GGA,\cite{PhysRevB.107.165111,Pittalis2017} which also recovers properly the result of the Becke-Roussel collinear exchange functional~\cite{PhysRevA.39.3761} for closed-shell systems.

\subsection{Noncollinear Perdew-Zunger SIC}

Based on Eq. (\ref{eq:self_interaction}), we can propose a generalization of the PZ-SIC to the noncollinear case. Let us first start by reviewing briefly the collinear case. The idea behind the PZ-SIC consists in removing all the single-electron self-interaction errors for a given density functional approximation. This leads to the energy functional
\begin{eqnarray}
 E_\xc^{\mathrm{SIC}} &=& E_\xc^{\mathrm{DFT}}[n_\uparrow,n_\downarrow] \nonumber\\
 &-& \sum_{\sigma=\{\uparrow,\downarrow\}} \sum_{i}f_{i,\sigma} \left(E_{\mathrm{H}}[n_{i\sigma}] + E_\xc^{\mathrm{DFT}}[n_{i\sigma},0]\right).
 \label{eq:pz_col}
\end{eqnarray}
In this expression, $n_\uparrow$  and $n_\downarrow$ refer respectively to the up and down channels of the total electronic density, and the $f_{i,\sigma}$ are occupation
numbers. For each Kohn-Sham orbital $\varphi_i$ one needs to compute the corresponding Hartree and exchange-correlation energy from its individual density $n_{i\sigma}$ and subtract it from the energy computed from the total density.

This above expression is intrinsically limited to the collinear case, but can be easily generalized to the noncollinear case.
Indeed, in the latter case the exchange-correlation functional is not a functional of the density in the two spin channels ($E_\xc[n_\uparrow,n_\downarrow]$) but a functional of the total density $n$ and the local magnetization $\mathbf{m}$.
This immediately suggests generalizing Eq. (\ref{eq:pz_col}) to the noncollinear case as
\begin{equation}
 E_\xc^{\mathrm{SIC}} = E_\xc^{\mathrm{DFT}}[n,\mathbf{m}] -  \sum_{i}f_i \left(E_{\mathrm{H}}[n_{i}] + E_\xc^{\mathrm{DFT}}[n_{i},\mathbf{m}_i]\right)\,.
 \label{eq:pz_noncol}
\end{equation}
This correction removes the self-interaction of each orbital $\varphi_i$ as in the collinear case.

In practice, the noncollinear PZ-SIC scheme can be challenging to implement. First of all, it requires finding the local effective potential originating from this orbital-dependent scheme, unless one wants to resort to using a generalized Kohn-Sham scheme that allows for orbital-dependent potentials.\cite{Seidl1996} Finding this local multiplicative potential is usually achieved by solving the OEP equation,\cite{Talman1976,RevModPhys.80.3} or some simplified version of it like the KLI approximation.\cite{Krieger1992}

A more subtle complexity comes from the fact that different orbitals can produce the same density. For a typical density functional approximation like LSDA, this is not a problem. However, this becomes a well-known problem with PZ-SIC, whose results depend on the orbitals and hence vary under a unitary transformation of the orbitals,\cite{PhysRevA.55.1765,PhysRevB.28.5992,10.1063/1.446959,10.1063/1.448266,10.1063/1.454104} unless one minimizes explicitly over all possible unitary transformations,\cite{RevModPhys.80.3,doi:10.1021/acs.jctc.5b00806} or use specific orbitals that make the SIC a true density functional.\cite{10.1063/1.4869581}  We will briefly discuss this point below with numerical examples.


Finally, let us comment on an important difference between the collinear case and the noncollinear case, which concerns the practical solution of the KLI equations to get to an approximate solution to the full OEP equation. When solving these equations, the potential is defined up to a constant, which is fixed by imposing for isolated systems that $v_{\mathrm{xc},\sigma}\to0$ for $r$ going to infinity.\cite{Krieger1992} This leads to a different constant for the up and down potentials in the collinear case. However, in the noncollinear case, we end up with a single constant, as we have a $2\times2$ matrix in spin space for the potential. As a consequence, for an open-shell system without SOC, for which we can compare directly the collinear and noncollinear results, the potentials for the majority spin are very similar, but in the minority spin channel they may be different.

\subsection{Noncollinear averaged density SIC}

While the PZ-SIC is known to produce very good results, it is also known to be numerically expensive to evaluate, as one needs to solve one Poisson equation and compute the exchange-correlation energy for each occupied Kohn-Sham state, and one further needs to solve the OEP equations to obtain a local multiplicative potential needed to perform Kohn-Sham SDFT calculations. This is why several simplified methods have been proposed. Among them, the most effective method is probably the AD-SIC, which, a bit surprisingly given its simplicity, can produce excellent results for atoms compared to PZ-SIC. The motivation of this method is that if all orbitals have a similar localization, we can replace their density in Eq. (\ref{eq:pz_col}) by their averaged density.\cite{RevModPhys.80.3} This is particularly suited for calculations with identical atoms and pseudopotential-based simulations as orbitals are similar in these cases. However, AD-SIC suffers from a size-consistency problem as it is explicitly based on the number of electrons,\cite{RevModPhys.80.3} which makes it unsuitable for extended systems. In this section, we show how to generalize  the AD-SIC to the noncollinear case.

In the collinear case, the AD-SIC is obtained by replacing in Eq. (\ref{eq:pz_col}) the orbital and spin-resolved density $n_{i\sigma}$ by the average spin-resolved density $n_{\sigma}/N_{\sigma}$, where $N_\sigma = \int d\mathbf{r}n_{\sigma}(\mathbf{r})$ is the number of electrons in the spin channel $\sigma$.
This directly leads to the collinear AD-SIC energy functional:
\begin{eqnarray}
 E_\xc^{\mathrm{AD-SIC}} &=&  E_\xc^{\mathrm{DFT}}[n_\uparrow,n_\downarrow] - \sum_{\sigma=\{\uparrow,\downarrow\}} N_\sigma\Big(E_{\mathrm{H}}[n_{\sigma}/N_\sigma] \nonumber\\
&+& E_\xc^{\mathrm{DFT}}[n_{\sigma}/N_\sigma,0]\Big)\,.
 \label{eq:AD-SIC_col}
\end{eqnarray}

Following this logic, one could be tempted to average not the up and down densities of collinear SDFT, but the full spin-density matrix of non-collinear SDFT, or equivalently the local charge and magnetization densities. Inserting this into Eq. (\ref{eq:pz_noncol}), one would directly obtain
\begin{eqnarray}
 E_\xc^{\mathrm{AD-SIC}} &=& E_\xc^{\mathrm{DFT}}[n,\mathbf{m}] -  N \Big(E_{\mathrm{H}}[n/N] \nonumber\\
&+& E_\xc^{\mathrm{DFT}}[n/N,\mathbf{m}/N]\Big)\,.
 \label{eq:AD-SIC_noncol}
\end{eqnarray}
However, this choice does not produce the correct collinear limit.
In order to illustrate this, let us consider a Li atom in a uniform magnetic field aligned along the $z$ direction. In this case, the system has three electrons, two residing in the 1s level, and one in the 2s level.  It is straightforward to see that the one electron in the 1s (spin-channel $\alpha$) and the one in the 2s level have their orbital magnetization antialigned with the external magnetic field, while the second 1s electron (spin-channel $\beta$) has its orbital magnetization aligned with the external magnetization. The densities corresponding to these states are denoted $n_{1s,\alpha}$, $n_{2s,\alpha}$ and $n_{1s,\beta}$.
Assuming that the approximate functional which we want corrected with AD-SIC fulfills the requirements mentioned in the introduction [SU(2) gauge invariance, and the same energy for a single orbital density with $m_z>0$ in the noncollinear case and for the same density in the up channel for the collinear functional] we can treat the same Li atom as a collinear electronic system with a static magnetic field along the $z$ axis.

Let us now compute the collinear and noncollinear AD-SIC corrections for this Li atom.
The AD-SIC for the collinear-spin case, Eq. (\ref{eq:AD-SIC_col}), is
\begin{eqnarray}
\Delta E^{\rm AD-SIC - col.} &=& -2E_{\rm H}[\frac{n_{1s, \alpha}+n_{2s,\alpha}}{2}] -E_{\rm H}[n_{1s, \beta}] \nonumber\\
&-&  2E_{\rm xc}[\frac{n_{1s, \alpha}+n_{2s,\alpha}}{2}, 0] -E_{\rm xc}[n_{1s,\beta}, 0]. \nonumber\\
\end{eqnarray}
If we use the proposed averaged density SIC as in Eq. (\ref{eq:AD-SIC_noncol}), we get
\begin{eqnarray}
\Delta E_\xc^{\mathrm{AD-SIC}} = -  3 \Big(E_{\mathrm{H}}[\frac{n_{1s,\alpha}+n_{2s,\alpha}+n_{1s,\beta}}{3}] \nonumber\\
 + E_\xc^{\mathrm{DFT}}[\frac{n_{1s,\alpha}+n_{2s,\alpha}+n_{1s,\beta}}{3}, \frac{\mathbf{m}_{1s,\alpha}+\mathbf{m}_{2s,\alpha}+\mathbf{m}_{1s,\beta}}{3}]\Big).\nonumber\\
\end{eqnarray}
Clearly, this expression will not lead to the desired collinear limit, as seen directly from the Hartree term.
However, it is possible to recover the collinear limit using the same logic as proposed originally by K\"ubler \textit{et al.}\cite{kubler1988density} for treating LSDA with noncollinear spin. By diagonalizing first the spin-density matrix, we obtain two densities, $n_\uparrow$ and $n_\downarrow$, which we can average by normalizing them by their integrals (thus defining the number of ``up'' and ``down'' electrons in the frame defined by the local magnetization). Similarly to the LSDA case, the potential is computed in the local frame and independently for the up and down channels, and then rotated back to the global frame using the total magnetisation. This procedure will produce the collinear limit expected in the above Li atom example.

The direct consequence of this procedure is that both the LSDA energy/potential and the AD-SIC corrections are evaluated in the same frame, which makes this approach
consistent and also invariant under local and global SU(2) rotations.
However, the price to pay is that the exchange-correlation magnetic field originating from the AD-SIC correction term is aligned with the local magnetisation, meaning that no exchange-correlation torque is produced by the correction scheme.

\section{Numerical results}
\label{sec:results}

We have implemented the above equations in the real-space code Octopus~\cite{Octopus_paper_2019} in order to perform tests. For the case of PZ-SIC, we only computed the solution of the OEP equations at the KLI level, using the explicit solution for noncollinear spin proposed in our recent work (see supplementary information of Ref. \onlinecite{PhysRevB.107.165111}).

\subsection{Isolated Xe atom}

\begin{figure}
  \begin{center}
    \includegraphics[width=0.9\columnwidth]{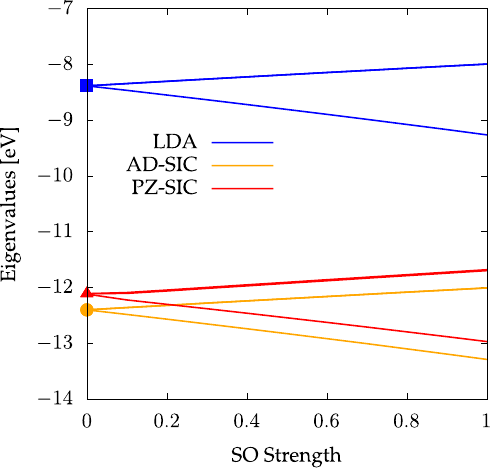}
    \caption{\label{fig:Xe_scaling} Splitting of 5p levels of Xe due to SOC versus the spin–orbit strength computed for LDA (blue curves), LDA+AD-SIC (orange curves), and LDA+PZ-SIC (red curves). The symbols (square, circle, and triangle) indicate the results obtained for the corresponding spin-unpolarized calculations.}
  \end{center}
\end{figure}

In order to investigate the interplay between SIC and SOC, as well as numerical and theoretical problems related to the various schemes, we first consider the case of an isolated Xe atom. We use a grid spacing of 0.30 Bohr, employing norm-conserving fully relativistic Hartwigsen-Goedecker-Hutter (HGH) pseudo-potentials.\cite{PhysRevB.58.3641} The simulation box is taken as a sphere of radius 12 Bohr centered at the atomic center.
In Fig.~\ref{fig:Xe_scaling} we show the splitting of the 5p electronic levels into 5p$_{1/2}$ and 5p$_{3/2}$ levels for LSDA, LSDA+AD-SIC, and LSDA+PZ-SIC. In all cases the collinear limit is correctly recovered for PZ-SIC and AD-SIC.
We found that the inclusion of the SIC does not change how SOC affects the energy levels, and the degeneracy of the energy levels is properly described by our corrections. As visible from the figure, we nicely recover the collinear limit, indicated by the symbols in Fig.~\ref{fig:Xe_scaling}.
We also checked that in the case of vanishing SOC strength, using a small magnetic field along $x$, $y$, or $z$ directions produces identical results, as expected from the SU(2) invariance of our proposed formulation. However, we note that for more complicated molecules, the collinear limit is not always recovered, see below.

Let us now comment on the dependence on a unitary transformation of the orbitals used in the evaluation of Eq. (\ref{eq:pz_col}) and Eq. (\ref{eq:pz_noncol}). In order to reveal this, we define a new set of orbitals $\{\tilde{\varphi}_i\}$ such that
\begin{equation}
 \tilde{\varphi}_i(\mathbf{r}) = \sum_j U_{ij}\varphi_j(\mathbf{r})\,,
\end{equation}
where $U$ is a unitary matrix. The two sets of orbitals, $\{\varphi_j\}$ and $\{\tilde{\varphi}_i\}$, have the same density, but their contribution to their PZ-SIC energy is different. To illustrate this we consider here three cases: i) the ``minimizing'' orbitals obtained directly from the solution of the Kohn-Sham equations, ii) the result of the so-called subspace diagonalization procedure in which the unitary matrix is found by diagonalizing the Hamiltonian matrix in the subspace of the ``minimizing'' orbitals, iii) the localization method known as the SCDM method\cite{damle2015compressed} that produces Wannier functions.
\begin{table}
\caption{\label{tab:xe} Total energy $E_{\rm tot}$ and ionization potential $I_p$, in Hartree, for the collinear and collinear cases using different orbitals, as explained in the main text. }
\begin{ruledtabular}
\begin{tabular}{lcccc}
 &  \multicolumn{2}{c}{Collinear case} &  \multicolumn{2}{c}{Noncollinear case} \\
 & Minimizing  & SCDM    & Minimizing   & SCDM \\
\hline\\
 $E_{\rm tot}$ & -15.5938 & -15.6492 & -15.5938 & -15.6499 \\
 $I_p$         & 0.4449   & 0.4674  & 0.4449 & 0.4695 \\
\end{tabular}
\end{ruledtabular}
\end{table}
In Table \ref{tab:xe} we report the total energy and ionization potential of Xe for the first and the last approach for both the collinear and the noncollinear case. We find no difference between the ``minimizing orbitals'' and the ones obtained by subspace diagonalization. As expected, more localized orbitals produce a lower total energy and a higher ionization potential.

Overall, it is apparent from these results that our noncollinear functional suffers from the same problems as the collinear formulation. One solution would be to implement a minimization of the PZ-SIC energy correction with respect to the unitary transformation $U$, which we defer to some future work. In the following, unless specified explicitly, orbitals from the subspace diagonalization are always employed.

\subsection{Diatomic closed-shell systems}

We continue analyzing the effect of our proposed functional on small closed-shell molecules for which SOC is known to be important for their electronic structure. It is known that SOC plays an important role on the bond length of closed-shell dimers, as well as their harmonic frequency and their dissociation energy.\cite{van1996zero} However, the choice of the functional is also important for these properties,\cite{van1996zero} and we expect the SIC to be relevant for improving the theoretical modelling of these molecules.

\begin{figure}
  \begin{center}
    \includegraphics[width=0.9\columnwidth]{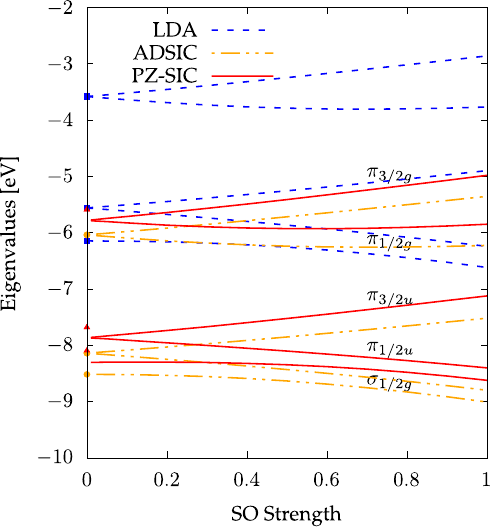}
    \caption{\label{fig:Bi2_scaling} Eigenvalues of the highest bonding ($\sigma_{1/2g}$, $\pi_{1/2u}$, and $\pi_{3/2u}$) and lowest antibonding ($\pi_{1/2g}$ and $\pi_{3/2g}$) molecular orbitals of the bismuth dimer as a function of the
SOC strength computed for LDA (blue curves), LDA+AD-SIC (orange curves), and LDA+PZ-SIC (red curves). The dots indicate the results obtained for spin-unpolarized calculations.}
  \end{center}
\end{figure}

We start by considering the Bi$_2$ molecule. We performed calculations at the experimental bond length\cite{huber1979constants} of 2.661 {\AA} for LSDA, LSDA+AD-SIC, and LSDA+PZ-SIC. We used a grid spacing of 0.30 Bohr, employing norm-conserving fully relativistic Hartwigsen-Goedecker-Hutter (HGH) pseudo-potentials.\cite{PhysRevB.58.3641} The simulation box was obtained from the union of two spheres of radii 12 Bohr centered on each atoms.
As shown in Fig.~\ref{fig:Bi2_scaling}, the inclusion of the SIC does not change how SOC affects the energy levels of the molecules, and the degeneracy of the energy levels is properly described by our corrections.

As in the case of Xe, the AD-SIC properly recovers the collinear limit, while we found that the PZ-SIC becomes unstable when the SOC strength is set to zero. Indeed, in this case Bi$_2$ is non-magnetic, and hence any local SU(2) rotation of the spins associated with a given orbital leaves the energy unchanged but changes the potential. In order to get a converged ground state in absence of SOC, we apply a tiny magnetic field. Unlike the case of Xe, we found here two possible solutions. Aligning the external magnetic field along the molecular axis, we get the limit of vanishing SOC strength. Aligning the magnetic field perpendicular to the molecular axis, we get the same eigenvalues as in the collinear calculation. This is analyzed more in detail in Appendix \ref{app:vanishing_soc_bi2}.

We also performed similar simulations for other diatomic molecules using their experimental geometry, see Table \ref{tab:table_ip}. For all molecules we employ a grid spacing of 0.3 Bohr, a radius for atom-center spheres of 12 Bohr, except for Au for which we included semi-core states and used a grid spacing of 0.25 Bohr.
\begin{table*}
\caption{\label{tab:table_ip} Ionization potentials, in eV, of diatomic systems using their experimental geometry, including SOC,  for different energy functionals.}
\begin{ruledtabular}
\begin{tabular}{lccccccc}
            & Bi$_2$    & Au$_2$    & I$_2$     & HI        & IF        & PbO       & TlF\\
\hline\\
Exp.         & 7.3\footnote{Ref.~\onlinecite{gingerich1974gaseous}} & 9.5\footnote{Ref.~\onlinecite{stearns1973mass}} & 9.307\footnote{Ref.~\onlinecite{cockett1996zero}} & 10.386\footnote{Ref.~\onlinecite{eland1977photoionization}} & 10.54\footnote{Ref.~\onlinecite{colbourn1978he}} & 9.4\footnote{Ref.~\onlinecite{makarov1993use}} & 10.52\footnote{Ref.~\onlinecite{dehmer1973photoelectron}}\\
LSDA         & 4.898     & 6.104     & 6.062     & 6.627     & 6.549     & 6.373     & 5.959 \\
LSDA+AD-SIC   & 7.773     & 9.481     & 8.651     & 10.294    & 10.015    &  10.190   & 10.614\\
LSDA+PZ-SIC  & 7.120     & 9.324     & 8.252     & 10.028    & 9.481     &  10.372   & 11.778\\
\end{tabular}
\end{ruledtabular}
\end{table*}
Overall, we find that the inclusion of the SIC drastically improves the agreement with respect to the experiment for the ionization potential, as expected from the vast literature on collinear SIC.

We also investigated the polar diatomic molecules HI, IF, PbO, and TlF at their experimental geometry and compared the dipole moments for different level of description with the experimental values.
\begin{table}
\caption{\label{tab:table_dipole} Dipole moments, in Debye, of diatomic systems using their experimental geometry, including SOC,  for different energy functionals.}
\begin{ruledtabular}
\begin{tabular}{lcccc}
            & HI        & IF        & PbO   & TlF\\
\hline\\
Exp.\footnote{Ref.~\onlinecite{huber1979constants}} & 0.45 & 1.95 & 4.64 & 4.23\\
LSDA         & 0.451     & 1.371     & 4.310 & 4.457\\
LSDA+AD-SIC   & 0.534     & 2.512     & 6.235 & 6.590\\
LSDA+PZ-SIC  & 0.372     & 1.483     & 5.569 & 4.612\\
\end{tabular}
\end{ruledtabular}
\end{table}
Consistent with the collinear case,\cite{10.1063/5.0024776} we found that the dipole moment on average deviates more from the experimental value when using SIC than simply using noncollinear LSDA. Importantly, the limitations of the approximation of an averaged density used to get to AD-SIC appears more clearly on the dipole moments than on the ionization energy.
We also performed geometry relaxation. As found in the collinear case,\cite{PhysRevA.55.1765,johnson2019effect} we obtain that including SIC shortens the bonds, resulting here in underestimated bond lengths compared to the LSDA, the latter being in better agreement with experimental values.

\subsection{Magnetic cluster}

We now investigate the effect of SIC on the properties of small magnetic clusters by specifically considering the iron dimer, Fe$_2$.\cite{footnote}
Clusters of this type have been widely studied by means of LSDA, including SOC (see for instance Ref. \onlinecite{akturk2016bh} and references therein). Unless stated differently, SOC is included throughout.
In all calculations we employ a grid spacing of 0.15 Bohr, a radius for atom-center spheres of 12 Bohr, and we included the semi-core states for Fe atoms. A small Fermi-Dirac smearing of 10 meV for the occupations was also used.
The Fe-Fe distance was taken for the Fe dimer to be the experimental one of 2.02 {\AA}.\cite{PhysRevB.25.4412}

\begin{table}
\caption{\label{tab:table_magnetic} Electronic and magnetic properties of Fe$_2$ for different energy functionals. Ionization potential ($I_p$) is given in eV, and the total ($M$) and local magnetic moments ($|\mathbf{m}|$) are given in $\mu_B$ and are obtained by integrating the density on a sphere of radius 1.909 Bohr around the atoms. Exchange-only LSDA (LSDAx) results and also reported.}
\begin{ruledtabular}
\begin{tabular}{lccc}
            & $I_p$  & $M$  & $|\mathbf{m}|$ \\
\hline\\
LSDA         & 3.327 & 6.00 & 2.71 \\
LSDA+AD-SIC   & 7.854 & 6.00 & 2.69 \\
LSDA+PZ-SIC  & 6.843 & 6.00 & 2.59 \\
\hline
LSDAx        & 3.453 & 8.00 & 3.29 \\
LSDAx+AD-SIC  & 7.464 & 7.00 & 2.97 \\
LSDAx+PZ-SIC & 5.995 & 7.50 & 3.10 \\
Slater       & 6.760 & 6.00 & 2.96 \\
\end{tabular}
\end{ruledtabular}
\end{table}
In all cases that included both LSDA exchange and correlation energy, we found a total magnetic moment of 6$\mu_B$, in agreement with in prior works. We note that our LSDA value matches well the atomic magnetic moment reported in the pioneering work of Oda {\em et al.}\cite{PhysRevLett.80.3622} The fact that the atomic magnetic moments computed on a sphere around the atoms decrease indicates that for Fe$_2$, the SIC tends to push away the magnetization from the atomic center, while the increase of the ionization potential is consistent with an increased localization of the orbitals. This points toward a non-negligible contribution of itinerant electrons to the magnetic properties in this cluster.
We also computed the values for exchange-only LSDA, together with SIC corrections. The total magnetic moments are not properly predicted in these cases, demonstrating the key importance of correlations in order to obtain reliable magnetic structures.

\begin{figure*}
  \begin{center}
    \includegraphics[width=0.9\textwidth]{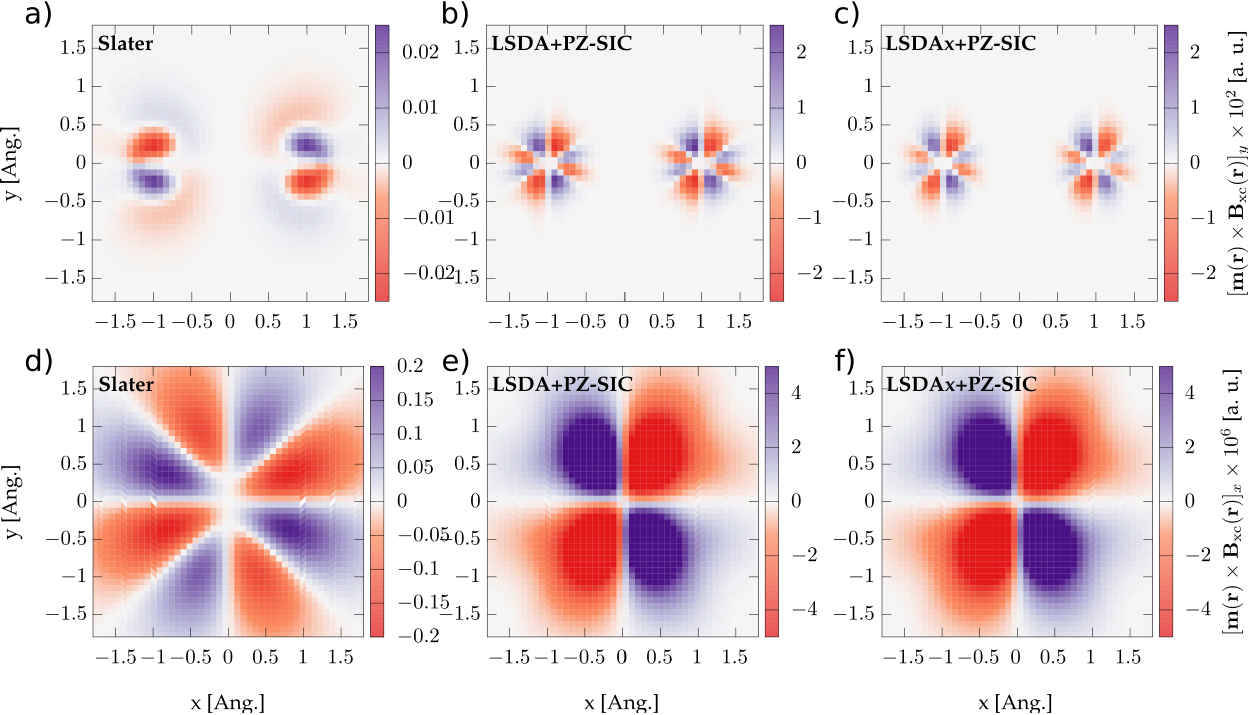}
    \caption{\label{fig:torque} Exchange torque for Fe$_2$. Top panels: The $y$ component of the local exchange torque $\tau(\mathbf{r})$ in the plane $y=0$, computed from a) Slater, b) LSDA with PZ-SIC, c) LSDAx with PZ-SIC. Bottom panels: the same as top panels, but showing $x$ component of the torque in the $z=0$ plane.}
  \end{center}
\end{figure*}

We finally turn our attention to the exchange-correlation torque $\tau(\mathbf{r})$, defined as
\begin{equation}
 \tau(\mathbf{r}) = \mathbf{m}(\mathbf{r})\times\mathbf{B}_{\mathrm{xc}}(\mathbf{r})\,,
\end{equation}
where $\mathbf{m}$ is the local magnetization density and $\mathbf{B}_{\mathrm{xc}}$ is the exchange-correlation magnetic field. We computed this quantity
using LSDA and LSDAx with PZ-SIC, and also with the Slater potential. As a reference here, we consider the Slater potential, which was shown to give reasonable results compared to the result of exact-exchange potential computed at the level of KLI.\cite{PhysRevB.107.165111} From our results (see Figs.~\ref{fig:torque}a and d) the Slater potential produces a small exchange-correlation torque around the atoms, where the symmetries of the system are clearly apparent.
Our results for PZ-SIC (Figs.~\ref{fig:torque}b, c, e, and f) show that PZ-SIC also produces a non-vanishing torque around the atoms. While it shows, as required by the zero-torque theorem, alternating positive and negative patterns that are also in accordance with the symmetries of the system, the overall shape and magnitude strongly differs from what is obtained from Slater potential.

Importantly, we want to stress here that like the energy, the torque obtained from PZ-SIC depends upon the unitary transformation of the orbitals. This quantity therefore needs to be analyzed with great care, and we aim in the future at implementing a minimization over unitary transformations in order to eliminate this ambiguity, similarly to prior efforts.\cite{doi:10.1021/acs.jctc.5b00806}

\section{Conclusions}
\label{sec:discussion}

To summarize, we presented how to extend some of the existing SIC approaches to the case of non-collinear spins.  We then analyzed numerically how these non-collinear SIC schemes behave for various closed-shell and magnetic systems. Overall, we found that our noncollinear schemes exhibit similar advantages and deficiencies as the collinear ones. The ionization energies are improved, but bond lengths are found to be worse than those obtained for LSDA. When the localization of individual orbitals is important, the AD-SIC performs poorly for observables that depend on local orbitals, like dipole moments or magnetic moments.

We further demonstrated that PZ-SIC for noncollinear spin can produce a non-negligible exchange-correlation torque around the magnetic atoms, but we found large differences in the magnitude and texture of the exchange-correlation torque compared to the result of the Slater potential.

Overall, our work opens the door to a better description of the electronic and magnetic properties of systems when noncollinear effects are important, but we note that some further work, including the computation of accurate benchmarks, is needed in order to get reliable results for the collinear and noncollinear PZ-SIC schemes.
Once such SIC schemes are fully established we expect them to become a useful tool for the description of materials with noncollinear magnetism.

\acknowledgments

C.A.U. is supported by DOE Grant No. DE-SC0019109.

\appendix
\section{Vanishing SOC limit in Bi$_2$}
\label{app:vanishing_soc_bi2}

\begin{figure}
  \begin{center}
    \includegraphics[width=0.9\columnwidth]{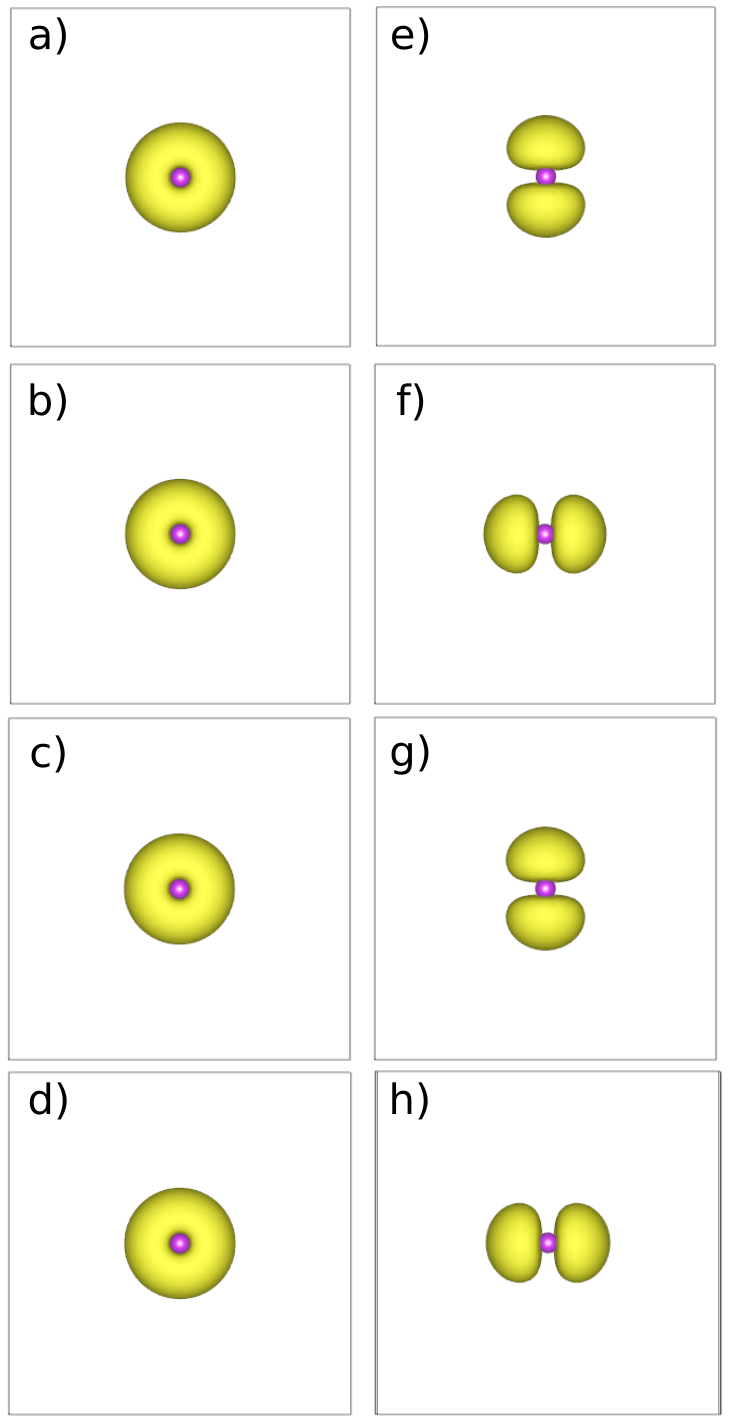}
    \caption{\label{fig:orbitals} Square modulus of the four highest occupied $\pi_u$ orbitals of Bi$_2$ obtained for a magnetic field aligned with the molecular axis a)-d); or perpendicular to it e)-h). The plane shown in the figures is the plane perpendicular to the molecular axis.}
  \end{center}
\end{figure}

In this section we investigate in more detail the case of Bi$_2$ without SOC using PZ-SIC, with a tiny magnetic field included. As explained in the main text, 
the dependence in the orbitals leads to different results for a magnetic field aligned with the molecular axis as opposed to saligned perpendicular to it.
In Fig.~\ref{fig:orbitals} we report the square modulus of the four highest occupied states of Bi$_2$ computed with PZ-SIC, corresponding to the $\pi_{1/2u}$ and $\pi_{3/2u}$ bounding orbitals, without SOC, and with a magnetic field aligned with the molecular axis or perpendicular to it.
While these orbitals produce the same charge density when summed over, their individual contributions for the PZ-SIC energy and potential is different, leading to a different ground state.

While these result might be surprising at first glance, the reported shapes are in fact the direct consequence of the symmetries of the system.
When the system has a magnetic field aligned with the molecular axis, it is clear that the system is invariant under any rotation along this axis. It is therefore not surprising to find radially symmetric wavefunctions in the panels a)-d).
On the contrary, when a tiny magnetic field is applied perpendicular to the molecular axis, the radial symmetry is broken, resulting in the splitting of the orbitals into two sets, one aligned with the magnetic field (e); g)) and one perpendicular to it (f); h)).  

The obtained wavefunctions therefore respect the symmetries of the system in the presence of a tiny magnetic field, and it is therefore expected that taking directly these orbitals to build the PZ-SIC energy (and the potential following from it) leads to some differences, even if the magnetic field itself has a negligible effect on the charge density.
Importantly, the change in the orbitals leads to a large change of 35mH in the total energy (the molecular-axis-aligned magnetic field giving the lowest energy), while the magnetic field itself only causes a splitting of these four energy levels by 15$\mu$H.

\bibliography{bibliography}

\begin{thebibliography}{48}
\bibitem [{foo()}]{footnote}%
  \BibitemOpen
  \href@noop {} {}\bibinfo {note} {We also considered Fe$_3$, which led to
  similar conclusions as Fe$_2$, so the results are not shown
  here.}\BibitemShut {Stop}%
\end{thebibliography}
\end{document}